\documentclass[11pt]{article}
\usepackage{amsmath,amsthm,latexsym,amssymb,amsfonts,epsfig}

\makeatletter
\@addtoreset{equation}{section}
\makeatother

\newtheorem{Theorem}{Theorem}[section]
\newtheorem{Definition}{Definition}[section]

\def\be{\begin{equation}}
\def\ee{\end{equation}}
\def\ba{\begin{eqnarray}}
\def\ea{\end{eqnarray}}

\def\a{{\cal A}}
\def\ab{\overline{\a}}

\def\Nl{{\mathchoice
{\setbox0=\hbox{$\displaystyle\rm N$}\hbox{\hbox to0pt
{\kern0.4\wd0\vrule height0.9\ht0\hss}\box0}}
{\setbox0=\hbox{$\textstyle\rm N$}\hbox{\hbox to0pt
{\kern0.4\wd0\vrule height0.9\ht0\hss}\box0}}
{\setbox0=\hbox{$\scriptstyle\rm N$}\hbox{\hbox to0pt
{\kern0.4\wd0\vrule height0.9\ht0\hss}\box0}}
{\setbox0=\hbox{$\scriptscriptstyle\rm N$}\hbox{\hbox to0pt
{\kern0.4\wd0\vrule height0.9\ht0\hss}\box0}}}}
\def\Zl{{\mathchoice
{\setbox0=\hbox{$\displaystyle\rm Z$}\hbox{\hbox to0pt
{\kern0.4\wd0\vrule height0.9\ht0\hss}\box0}}
{\setbox0=\hbox{$\textstyle\rm Z$}\hbox{\hbox to0pt
{\kern0.4\wd0\vrule height0.9\ht0\hss}\box0}}
{\setbox0=\hbox{$\scriptstyle\rm Z$}\hbox{\hbox to0pt
{\kern0.4\wd0\vrule height0.9\ht0\hss}\box0}}
{\setbox0=\hbox{$\scriptscriptstyle\rm Z$}\hbox{\hbox to0pt
{\kern0.4\wd0\vrule height0.9\ht0\hss}\box0}}}}
\def\Ql{{\mathchoice
{\setbox0=\hbox{$\displaystyle\rm Q$}\hbox{\hbox to0pt
{\kern0.4\wd0\vrule height0.9\ht0\hss}\box0}}
{\setbox0=\hbox{$\textstyle\rm Q$}\hbox{\hbox to0pt
{\kern0.4\wd0\vrule height0.9\ht0\hss}\box0}}
{\setbox0=\hbox{$\scriptstyle\rm Q$}\hbox{\hbox to0pt
{\kern0.4\wd0\vrule height0.9\ht0\hss}\box0}}
{\setbox0=\hbox{$\scriptscriptstyle\rm Q$}\hbox{\hbox to0pt
{\kern0.4\wd0\vrule height0.9\ht0\hss}\box0}}}}
\def\Rl{{\mathchoice
{\setbox0=\hbox{$\displaystyle\rm R$}\hbox{\hbox to0pt
{\kern0.4\wd0\vrule height0.9\ht0\hss}\box0}}
{\setbox0=\hbox{$\textstyle\rm R$}\hbox{\hbox to0pt
{\kern0.4\wd0\vrule height0.9\ht0\hss}\box0}}
{\setbox0=\hbox{$\scriptstyle\rm R$}\hbox{\hbox to0pt
{\kern0.4\wd0\vrule height0.9\ht0\hss}\box0}}
{\setbox0=\hbox{$\scriptscriptstyle\rm R$}\hbox{\hbox to0pt
{\kern0.4\wd0\vrule height0.9\ht0\hss}\box0}}}}
\def\Cl{{\mathchoice
{\setbox0=\hbox{$\displaystyle\rm C$}\hbox{\hbox to0pt
{\kern0.4\wd0\vrule height0.9\ht0\hss}\box0}}
{\setbox0=\hbox{$\textstyle\rm C$}\hbox{\hbox to0pt
{\kern0.4\wd0\vrule height0.9\ht0\hss}\box0}}
{\setbox0=\hbox{$\scriptstyle\rm C$}\hbox{\hbox to0pt
{\kern0.4\wd0\vrule height0.9\ht0\hss}\box0}}
{\setbox0=\hbox{$\scriptscriptstyle\rm C$}\hbox{\hbox to0pt
{\kern0.4\wd0\vrule height0.9\ht0\hss}\box0}}}}
\def\Hl{{\mathchoice
{\setbox0=\hbox{$\displaystyle\rm H$}\hbox{\hbox to0pt
{\kern0.4\wd0\vrule height0.9\ht0\hss}\box0}}
{\setbox0=\hbox{$\textstyle\rm H$}\hbox{\hbox to0pt
{\kern0.4\wd0\vrule height0.9\ht0\hss}\box0}}
{\setbox0=\hbox{$\scriptstyle\rm H$}\hbox{\hbox to0pt
{\kern0.4\wd0\vrule height0.9\ht0\hss}\box0}}
{\setbox0=\hbox{$\scriptscriptstyle\rm H$}\hbox{\hbox to0pt
{\kern0.4\wd0\vrule height0.9\ht0\hss}\box0}}}}
\def\Ol{{\mathchoice
{\setbox0=\hbox{$\displaystyle\rm O$}\hbox{\hbox to0pt
{\kern0.4\wd0\vrule height0.9\ht0\hss}\box0}}
{\setbox0=\hbox{$\textstyle\rm O$}\hbox{\hbox to0pt
{\kern0.4\wd0\vrule height0.9\ht0\hss}\box0}}
{\setbox0=\hbox{$\scriptstyle\rm O$}\hbox{\hbox to0pt
{\kern0.4\wd0\vrule height0.9\ht0\hss}\box0}}
{\setbox0=\hbox{$\scriptscriptstyle\rm O$}\hbox{\hbox to0pt
{\kern0.4\wd0\vrule height0.9\ht0\hss}\box0}}}}
\DeclareMathOperator{\MC}{\boldsymbol{\mathsf{M}}}
\DeclareMathOperator{\MCO}{\boldsymbol{\widehat{\mathsf{M}}}}

\DeclareMathOperator{\MCW}{\rm Master\;\;Constraint}
\DeclareMathOperator{\MCOW}{\rm Master\;\;Constraint\;\;
Operator}
\DeclareMathOperator{\MCPW}{\rm Master\;\;Constraint\;\;Programme}

\title{Testing the\\ $\MCPW$\\ for Loop Quantum Gravity\\
V. Interacting Field Theories}
\author{B.
Dittrich\thanks{dittrich@aei.mpg.de, bdittrich@perimeterinstitute.ca},
T. Thiemann\thanks{thiemann@aei.mpg.de, tthiemann@perimeterinstitute.ca}\\
\\
Albert Einstein Institut, MPI f. Gravitationsphysik\\
Am M\"uhlenberg 1, 14476 Potsdam, Germany\\
\\
and\\
\\
Perimeter Institute for Theoretical Physics \\
31 Caroline Street North, Waterloo, ON N2L 2Y5, Canada}
\date{{\small Preprint AEI-2004-120}}
\begin{document}

\maketitle

\begin{abstract}
This is the final fifth paper in our series of five in which we test the 
Master Constraint Programme for solving the Hamiltonian constraint in Loop 
Quantum Gravity. Here we consider interacting quantum field theories, 
specifically we consider the non -- Abelean Gauss constraints of 
Einstein -- Yang -- Mills theory and 2+1 gravity. Interestingly, while 
Yang -- Mills theory in 4D is not yet rigorously defined as an ordinary
(Wightman) quantum field theory on Minkowski space, in background 
independent quantum field theories such as Loop Quantum Gravity (LQG) this 
might become possible by working in a new, background independent 
representation. 

While for the Gauss constraint the Master constraint can be solved 
explicitly, for the 2+1 theory we are only able to rigorously define the 
Master Constraint Operator. We show that the, by other methods 
known, physical Hilbert is contained in the kernel of the Master 
Constraint, however, to systematically derive it by only using 
spectral methods is as complicated as for 3+1 gravity and 
we therefore leave the complete analysis for 3+1 gravity.
 \end{abstract}

\newpage

\tableofcontents

\section{Introduction}
\label{s1}

We continuue the test of the Master Constraint Programme 
\cite{7.0} for Loop Quantum Gravity (LQG) \cite{1.1,7.2,7.3}
which we started in the companion papers \cite{I,II,III,IV}.
The Master Constraint Programme is a new idea to improve on 
the current situation with the Hamiltonian constraint operator for LQG
\cite{7.1}. In short, progress on the solution of the Hamiltonian 
constraint has been slow because of a technical reason: the Hamiltonian 
constraints 
themselves are not spatially diffeomorphism invariant. This means that one 
cannot first solve the spatial diffeomorphism constraints and then 
the Hamiltonian constraints because the latter do
 not preserve the 
space of solutions to the spatial diffeomorphism constraint
\cite{8.2}. On the other hand, the space of solutions  
to the spatial diffeomorphism constraint \cite{8.2} is relatively easy to 
construct starting from the spatially diffeomorphism invariant 
representations on which LQG is based \cite{7.4} which are therefore 
very natural to use and, moreover, essentially unique. Therefore one would 
really like to keep these structures. The Master Constraint 
Programme removes that technical obstacle by replacing the Hamiltonian 
constraints by a single Master Constraint which is a spatially 
diffeomorphism invariant integral of squares of the individual 
Hamiltonian constraints which encodes all the necessary information about 
the constraint surface and the associated invariants. See e.g. 
\cite{7.0,I} for a full discussion of 
these issues. Notice that the idea of squaring constraints is not new,
see e.g. \cite{2.1}, however, our concrete implementation is new and also 
the Direct Integral Decomposition (DID) method for solving them, see 
\cite{7.0,I} for all the details. 

The Master Constraint for four 
dimensional General Relativity will appear in \cite{8.1} but before we 
test its semiclassical limit, e.g. using the methods of \cite{8.3,8.5} 
and try to solve it by DID methods we want to test the programme in the  
series of papers \cite{I,II,III,IV}. 
In the previous papers we focussed on finite dimensional
systems of various degrees of complexity and free field theories.
In this paper we consider interacting quantum field theories,
in particular in four dimensions and 2+1 gravity which is interacting 
before one solves the constraint (the reduced phase space is of course 
finite dimensional). This might be a surprise given the fact 
that no interacting Wightman quantum field theories have been constructed 
so far in four dimensions. The resolution of the puzzle is that instead of 
working on a Minkowski background we couple gravity to matter and 
therefore arrive at a background independent quantum field theory.
That these display a better ultraviolet behaviour than background 
dependent quantum field theories, at least for a large class of operators 
was already shown in \cite{7.1} and we will see that the same mechanism 
is at work here and makes the associated Master Constraint Operators 
well defined. This does not mean, however, that we are done and have a 
consistent quantum theory of both matter and geometry because only
1. after having solved the Master Constraint for {\it all} constraints
(here we only consider the matter Gauss constraints but not, in 
particular, the Hamiltonian constraints) of gravity coupled to the 
standard model and 2. after having shown that the solution theory is 
consistent with experiment (reduces to QFT on (curved) backgrounds in the
semiclassical limit of low geometry fluctuations) do we have a viable 
candidate theory. It is precisely the purpose of the Master Constraint 
Programme to complete this task and the present series of papers serves 
the purpose to make sure that the Master Constraint Programme reproduces 
the established results in solvable cases.

\section{Infinite Number of Non-Abelean First Class Constraints 
Non -- Polynomial in the Momenta with Structure Constants}
\label{s7}

In a previous article \cite{IV} we discussed a background dependent 
quantum field
theory, specifically free Maxwell theory. Even in this free theory the 
definition of the $\MCW$
had to involve a non-trivial integral kernel $C$ which depended on the 
background metric in oder that the $\MCOW$ was densely defined. In this 
section we discuss two background independent theories, namely 
Einstein -- Yang -- Mills theory and pure Einstein gravity. Notice that 
these theories are no longer free, both the Yang -- Mills and 
gravitational
degrees of freedom are self-interacting and interacting with each other.
One therefore expects severe ultraviolet divergence problems, the more as 
we are not allowed to use a background metric in order to regulate 
the $\MCOW$. Nevertheless it turns out that {\bf precisely because of the 
background independence the ultraviolet problems can be overcome, 
background independent theories regulate themselves!} The way this works 
is identical to the mechanism discovered in \cite{7.1} for the definition 
of various Hamiltonian constraints in Loop Quantum Gravity (LQG) (see 
\cite{7.2,7.3} for an introduction). Hence we will construct 
a corresponding $\MCOW$ for the Gauss constraints of Einstein -- Yang -- 
Mills theory using LQG techniques.  

In what follows we consider canonical Yang -- Mills fields for a compact
gauge group $G$ with canonicaly conjugate pair 
$(\underline{A}_a^J,\underline{E}^a_J)$. Here $a,b,..=1,2,3$ are spatial 
tensor indices while $J,K,..=1,..,\dim(G)$ are Lie$(G)$ indices. 
Likewise for the gravitational sector we consider the canonical
pair $(A_a^j,E^a_j)$ where $j,k,..=1,2,3$ are $su(2)$ indices. From the
kinematical point of view the gravitational phase space is identical
with that for a Yang -- Mills theory with gauge group $SU(2)$. We assume 
both principal bundles to be trivial for simplicity. The non-vanishing
Poisson brackets are
\ba \label{7.1}
\{\underline{E}^a_J(x),\underline{A}_b^K(y)\} &=& g^2 \delta^a_b 
\delta_J^K \delta(x,y)      
\nonumber\\
\{E^a_j(x),A_b^k(y)\} &=& \kappa \delta^a_b 
\delta_j^k \delta(x,y)      
\ea
where $g^2$ and $\kappa=8\pi G_{Newton}$ denote the Yang -- Mills and 
gravitational coupling constant respectively. We are using units in which
both connections have dimension cm$^{-1}$ while the Yang -- Mills and 
gravitational electric fields respectively have dimension cm$^{-2}$ and 
cm$^0$ respectively. As a result the Feinstrukturkonstante 
$\alpha=\hbar g^2$ is dimensionless while $\ell_p^2=\hbar \kappa$ is the 
Planck area.

The kinematical phase space of the theory is subject to the 
(non -- Abelean) Gauss constraints
\ba \label{7.2}
\underline{G}_J &=& \partial_a \underline{E}^a_J+f_{JK}\;^L 
\underline{A}_a^K \underline{E}^a_L
\nonumber\\
G_j &=& \partial_a E^a_j+\epsilon_{jk}\;^l 
A_a^k E^a_l
\ea    
where in terms of a basis $\underline{\tau}_J,\tau_j$ for Lie$(G)$ and 
$su(2)$ respectively the structure constants are defined by 
$[\underline{\tau}_J,\underline{\tau}_K]=f_{JK}\;^L\tau_L$ and  
$[\tau_j,\tau_k]=\epsilon_{jk}\;^l\tau_l$ respectively. We normalize 
the anti Hermitean generators such that 
Tr$(\underline{\tau}_J \underline{\tau}_K)=-\delta_{JK}/2$ and    
Tr$(\tau_j \tau_k)=-\delta_{jk}/2$.
The phase space is subject to further spatial diffeomorphism and 
Hamiltonian constraints but these we reserve for a separate paper
\cite{8.1}.

We now review the kinematical Hilbert space of the theory which is 
actually selected by requiring spatial diffeomorphism invariance 
\cite{7.4}. It is easiest described in terms of so -- called spin 
(or charge, flavour, colour..)  network functions. We do this for a 
general compact group. 
\begin{Definition} \label{def7.1} ~~~~\\
i)\\
A spin network $s$ for a gauge theory with compact gauge group $G$ over a 
manifold $\sigma$ consists of a quadruple 
$s=(\gamma(s),\pi(s),m(s),n(s))$ consisting of an oriented (piecewise 
analytic) 
graph $\gamma(s)$ embedded into $\sigma$ as well as a labelling of each of
the edges $e$ of $\gamma(s)$ with a nontrivial irreducible representation 
$\pi_e(s)$ of $G$ and corresponding matrix element labels 
$m_e(s),n_e(s)=1,..,\dim(\pi_e(s))$. Here for each equivalence class 
of irreducible representations we have chosen once and for all an 
arbitrary representative.\\
ii)\\
A spin network function is simply the following complex valued function
on the space $\a$ of smooth connections over $\sigma$     
\be \label{7.5}
T_s(A):=\prod_{e\in E(\gamma(s))}\;[\sqrt{\dim(\pi_e(s))}\;
\{[\pi_e(s)](A(e))\}_{m_e(s) n_e(s)}]
\ee
where 
\be \label{7.6}
A(e):={\cal P} \exp_e(\int_e A^J \tau_J)
\ee
denotes the holonomy of $A$ along $e$ and $E(\gamma(s))$ is the set of 
edges of $\gamma(s)$.\\
iii)\\
The kinematical Hilbert space ${\cal H}_{Kin}$ is the closure of the 
finite linear span of spin network functions which define a basis.
Hence the kinematical inner product is given by
\be \label{7.7}
<T_s,T_{s'}>_{Kin}=\delta_{s,s'}
\ee
The holonomy operators $\widehat{A(e)}$ act by multiplication while 
the conjugate electric flux operators correspondinng to 
\be \label{7.7a}
E_J(S)=\int_S (\ast E_J)
\ee
act by differentiation whose details \cite{7.3} we will not need in what 
follows. Here $\ast E_J$ denotes the pseudo two -- form dual to the vector 
density $E^a_J$ of weight one.
\end{Definition}
One can show that ${\cal H}_{Kin}=L_2(\ab,d\mu_0)$ is a space of square 
integrable functions over a distributional extension $\ab$ of $\a$ with
respect to a probability Borel measure \cite{7.5,7.3} but this will not
be needed in what follows. For the Yang -- Mills and gravitational 
sector respectively we have one kinematical Hilbert space each 
corresponding to the gauge group under question and $SU(2)$ respectively 
and we will denote them as ${\cal H}_{Kin}^{YM}$ and ${\cal H}_{Kin}^{GR}$
respectively. The total kinematical Hilbert space for the 
Einstein -- Yang -- Mills theory is simply the tensor product
${\cal H}_{Kin}={\cal H}^{YM}_{Kin}\otimes {\cal H}^{GR}_{Kin}$. 
A convenient basis is given by the states $T_c \otimes T_s$ where 
$c,s$ respectively are spin networks for $G$, called colour networks in 
what follows, and for $SU(2)$ respectively. Please refer to 
\cite{7.3} and the sixth and eigth reference of \cite{7.1} for more 
details.

The functions (\ref{7.2}) are scalar densities of weight one with respect 
to spatial diffeomorphisms Diff$(\sigma)$ and they transform in the 
adjoint representation of $G$ and $SU(2)$ respectively. In \cite{7.3}
we showed that their smeared form admits a well-defined quantization as 
essentially self-adjoint operators on ${\cal H}_{Kin}$. More specifically,
let $\underline{\Lambda},\Lambda$ be a $Lie(G)-$ or $su(2)-$valued 
functions on $\sigma$ respectively (not 
necessarily smooth!) and let
\be \label{7.7b}
\underline{G}(\underline{\Lambda}):=\int_\sigma d^3x 
\underline{\Lambda}^J\underline{G}_J
\ee
and similarly for $G(\Lambda)$. Then 
\ba \label{7.8}  
\widehat{\underline{G}(\underline{\Lambda})}\;T_c\otimes T_s
&=& \{i\alpha \sum_{v\in V(\gamma(c)} \underline{\Lambda}^J(v)
{[}\sum_{e\in E(\gamma(c));b(e)=v} \underline{R}^e_J-
 \sum_{e\in E(\gamma(c));f(e)=v} \underline{L}^e_J]\;\;T_c\}\otimes T_s
\nonumber\\
&=:& \{i\alpha \sum_{v\in V(\gamma(c)} \underline{\Lambda}^J(v) 
\underline{X}^v_J\;\;T_c\}\otimes T_s
\nonumber\\
\widehat{G(\Lambda)}\;T_c\otimes T_s
&=& T_c\otimes\{i\ell_p^2 \sum_{v\in V(\gamma(s))} \Lambda^j(v)
{[}\sum_{e\in E(\gamma(s));b(e)=v} R^e_j-
 \sum_{e\in E(\gamma(s));f(e)=v} L^e_j]\;\; T_s\}
\nonumber\\
&=:& T_c\otimes\{i\ell_p^2\sum_{v\in V(\gamma(c))} \Lambda^j(v) X^v_j
\;\; T_s\}
\ea
Here $V(\gamma)$ denotes the set of vertices of a graph $\gamma$, 
$b(e),f(e)$ 
respectively denote beginning and final point of an edge $e$ and the 
operators $R^e_j,L^e_j$ respectively act as follows (the description is 
similar for the Yang -- Mills counterparts): We may view spin network 
states as so -- called cylindrical functions
\be \label{7.9} 
T_s(A)=f_{\gamma(s)}(\{A(e)\}_{e\in E(\gamma(s))})
\ee
where $f_{\gamma(s)}$ is a complex valued function on 
$SU(2)^{|E(\gamma(s))|}$. Then
\ba \label{7.10a}
(R^e_j T_s)(A) &:=&(\frac{d}{dt})_{t=0} 
f_{\gamma(s)}(\{e^{t\tau_j \delta_{ee'}}A(e')\}_{e'\in E(\gamma(s))})
\nonumber\\
(L^e_j T_s)(A) &:=&(\frac{d}{dt})_{t=0} 
f_{\gamma(s)}(\{A(e')e^{t\tau_j \delta_{ee'}}\}_{e'\in E(\gamma(s))})
\ea
This ends our review of the kinematical LQG decription of 
Einstein -- Yang -- Mills theory. In the next two subsections we will 
construct and solve the associated $\MCOW$ corresponding to both Gauss
constraints.

\subsection{Einstein -- Yang -- Mills Gauss Constraint}
\label{s7.1}

In \cite{7.1} it was shown that background independent theories offer the 
possibility to define ultraviolet finite operators for classical integrals
of scalar densities of weight one. The intuitive reason for this is as
follows: If $F=\int d^3x f(x)$ is the classical integral to be quantized 
then the density $n$ scalar $f(x)$ becomes an operator valued distribution 
of the form 
$\hat{f}(x)=\sum_\alpha (\delta(x,x_\alpha))^n \hat{f}_\alpha$ where 
the sum is over some index set depending on the theory and 
$f$. Here $\hat{f}_\alpha$ is an actual operator (not a distribution).
Notice that the cassical density weight of $f$ is correctly carried
by the $n-th$ power of the $\delta-$distribution but unless $n=1$ this 
is ill-defined and requires a point splitting regularization 
(``operator product expansion'') with subsequent renormalization.
This is actually the reason why in \cite{IV} we needed a trace 
class operator in order to split the points of the square 
$(\partial E)^2$ which is a scalar density of weight two. Splitting points 
requires a background metric which is not allowed in background 
independent theories, however, on the other hand in background independent
theories the density weight comes out to be {\bf always unity because of 
spatial diffeomorphism invariance!} Namely only integrals of scalar 
densities of weight one are spatially diffeomorphism invariant. 
Hence in background independent theories renormalization is not only 
not allowed but also not needed. The 
interested reader is referred to the second reference in \cite{1.1}
and to \cite{7.3} for more details. 

In keeping with this spirit we want to define a $\MCW$ which is the 
integral 
of a density of weight one, quadratic in the Gauss constraint and 
independent of a background metric. The simplest and most natural choice 
is 
\be \label{7.10}
\MC:=\int_\sigma \;d^3x\; 
\frac{\underline{G}_J\underline{G}_K \delta^{JK}}{\sqrt{\det(q)}}
\ee
Here the spatial metric $q_{ab}$ is defined via its inverse
$\det(q)q^{ab}=E^a_j E^b_k\delta^{jk}$. Notice that the overall density 
weight of (\ref{7.10}) equals unity indeed. In fact, (\ref{7.10}) is both 
$G-$invariant and Diff$(\sigma)-$invariant. The choice (\ref{7.10}) is 
natural because $\sqrt{\det(q)}$ is the simplest scalar density of weight 
one which at least classically is nowhere vanishing. We could not have 
proceeded
similarly with pure Maxwell theory because there the simplest scalar
density is $\partial\cdot E$ itself which would have resulted in 
$\MC=\int d^3x |\partial \cdot E|$. Not only would it be impossible to 
quantize 
this expression on Fock space, it is also not a classically differentiable 
function on the 
classical phase space and hence inacceptable. 
In non -- Abelean Yang -- Mills theory it is also possible to construct
the ``metric'' $Q_{ab}$ whose inverse is defined by $\det(Q) Q^{ab}
=\underline{E}^a_J \underline{E}^b_K \delta^{JK}$, however, the 
corresponding scalar density $\sqrt{\det(Q)}$ which does not vanish 
identically if $\dim(G)\ge 3$ is classically not constrained to be non -- 
vanishing everywhere. Hence the fact that we have coupled gravity cannot 
be avoided.

To quantize (\ref{7.10}) we will proceed similarly as in \cite{7.1}. Hence
we will be brief, referring the interested reader to the literature. 
The idea is to make use of the identity
\be \label{7.11}
1=\frac{[\det(e)]^2}{\det(q)}
\ee
where $e_a^j$ is the co -- triad defined up to $SU(2)$ 
transformations by $q_{ab}=\delta_{jk} e^j_a e^k_b$. 
Let ${\cal P}$ be a partition of $\sigma$ into mutually disjoint regions 
$R$. Then the integral (\ref{7.10}) is the limit as ${\cal P}\to\sigma$
of the corresponding Riemann sum
\ba \label{7.12}
\MC &=&\lim_{{\cal P}\to\sigma} \sum_{R\in {\cal P}}\;
\frac{\underline{G}_J(R)\underline{G}_K(R)\delta^{JK} e(R)^2}{V(R)^3}
\nonumber\\
\underline{G}_J(R) &=&\int_R d^3x \underline{G}_J(x)
\nonumber\\
e(R) &=&\int_R d^3x \det(e)(x)
\nonumber\\
V(R) &=&\int_R d^3x \sqrt{\det(q)}(x)
\ea
We now observe the further identity
\be \label{7.13}
e(R)=(\frac{1}{\kappa})^3\int_R \epsilon_{jkl} 
\{A^j(x),V(R)\}  \wedge
\{A^k(x),V(R)\}  \wedge
\{A^l(x),V(R)\}  
\ee
hence 
\be \label{7.14}
\frac{e(R)}{V(R)^{3/2}}=(\frac{2}{\kappa})^3
\int_R \epsilon_{jkl} 
\{A^j(x),V(R)^{1/2}\}  \wedge
\{A^k(x),V(R)^{1/2}\}  \wedge
\{A^l(x),V(R)^{1/2}\}  
=:(\frac{2}{\kappa})^3 \tilde{e}(R)
\ee
Thus (\ref{7.12}) becomes
\be \label{7.15}
\MC=\lim_{{\cal P}\to\sigma} (\frac{2}{\kappa})^6\sum_{R\in {\cal P}}\;
\delta^{JK}\underline{G}_J(R)\underline{G}_K(R)
\tilde{e}(R)^2
\ee
The reason for writing the $\MCW$ in this form is that all quantities 
involved in (\ref{7.15}) admit a well-defined quantization at finite 
partition $\cal P$: First of all $\underline{G}_J(R)=
\underline{G}(\underline{\Lambda})$ with $\underline{\Lambda}^K=
\chi_R\delta_J^K$ where $\chi_R$ is the characteristic function of 
$R$ admits a well-defined quantization according to (\ref{7.8}). 
Next the volume $V(R)$ of $R$ is well-defined as a positive essentially
self-adjoint operator on ${\cal H}_{Kin}$, its explicit action on spin 
network functions being given by
\be \label{7.16}
\hat{V}(R)T_s=\ell_p^3\sum_{v\in V(\gamma(s))\cap R} 
\sqrt{\frac{1}{48}|\sum_{e_1,e_2,e_3\in E(\gamma(s));
b(e_1)=b(e_2)=b(e_3)=v}\epsilon^{jkl}\epsilon(e_1,e_2,e_3)
R^{e_1}_j R^{e_2}_k R^{e_3}_l|}\;\;T_s
\ee
Here we have assumed that all edges are outgoing from a vertex (split
edges into two halves to do that) and $\epsilon(e_1,e_2,e_3)$ 
is the sign of the determinant of the matrix defined by the column 
vectors $\dot{e}_1(0),\dot{e}_2(0)\dot{e}_3(0)$ in this sequence where
$e_I(0)=b(e_I),\;I=1,2,3$. 

It remains to quantize $\tilde{e}(R)$ itself 
and to take the limit ${\cal P}\to\sigma$. To that effect, notice that the 
limit in (\ref{7.12}) is independent of the choice of the 
sequence of partitions ${\cal P}$. We may therefore without loss of 
generality assume that the partition is actually a triangulation 
consisting of tetrahedra $R$. In the limit ${\cal P}\to\sigma$ each $R$
can be described by a base point $v(R)$ and a right oriented triple 
of edges $e_I(R)\subset\partial R,\;I=1,2,3$, incident at and outgoing 
from $v(R)$. Then it is easy to see that in the limit ${\cal P}\to 
\sigma$
the expression $\tilde{e}(R)$ can be replaced by
\ba \label{7.17}
-8 e'(R) &=&-8\epsilon^{jkl} \epsilon^{JKL} 
\mbox{Tr}(\tau_j (A(e_J(R)))^{-1}\{A(e_J(R)),V(R)^{1/2}\})
\times\\
&&\times
\mbox{Tr}(\tau_k (A(e_K(R)))^{-1}\{A(e_K(R)),V(R)^{1/2}\}) \;\;
\mbox{Tr}(\tau_l (A(e_L(R)))^{-1}\{A(e_L(R)),V(R)^{1/2}\}) 
\nonumber
\ea
Expression (\ref{7.17}) can now be quantized by replacing Poisson brackets 
by commutators divided by $i\hbar$.  

Anticipating that the strong limit ${\cal P}\to\sigma$ exists as a 
symmetric operator we define
\be \label{7.17a}
\MCO=\lim_{{\cal P}\to\sigma} (\frac{4}{\kappa})^6\sum_{R\in {\cal P}}\;
\delta^{JK}\widehat{\underline{G}_J(R)}
\widehat{e'(R)}^2\widehat{\underline{G}_K(R)}
\ee
Notice that no ordering ambiguities 
arise because the gravitational and Yang -- Mills degrees of freedom 
commute with each other. It will be sufficient to define (\ref{7.17})
on the basis $T_c\otimes T_s$. 

In performing the limit a regularization ambiguity arises: The 
quantum limit when 
applied to $T_c\otimes T_s$ depends on the sequence ${\cal P}\to\sigma$
while the classical limit was independent of that choice. We will 
therefore 
proceed as in \cite{7.1} and choose a sequence for each $T_c\otimes T_s$
individually which is justified by the fact that classically the choice of 
the sequence does not make any difference. The result 
is the following: We notice first of all from (\ref{7.8}) that there is no 
contribution 
in (\ref{7.17}) from those tetrahedra $R$ which do not contain a vertex of 
$\gamma(c)$. Next, $[\hat{A}(e),\hat{V}(R)^{1/2}]$ evidently 
vanishes for those $R$ 
which do not contain a vertex of $\gamma(s)$ no matter how we choose
$e\subset \partial R$. Hence, for sufficiently fine ${\cal P}$ we can 
focus attention on those $R$ which contain a common vertex of 
$\gamma(s)$ and $\gamma(c)$. The idea is now to average over all 
triangulations. Let $n(v)$ be the gravitational valence of 
such a vertex then the detailed averaging performed in \cite{7.1} 
results
in the following final expression after taking the limit 
${\cal P}\to\sigma$ 
\ba \label{7.18}
\MCO T_c\otimes T_s &=&
\alpha^2(\frac{4}{\ell_p})^6\sum_{v\in V(\gamma(c))\cap 
V(\gamma(s))}\;
\delta^{JK}\underline{X}_J(v)
\widehat{e'(v)}^2\underline{X}_K(v) \;\;T_c\otimes T_s
\nonumber\\
\widehat{e'(v)}&=& \frac{1}{n(v)(n(v)-1)(n(v)-2)} \epsilon^{jkl}
\epsilon^{JKL} 
\sum_{e_1,e_2,e_3\in E(\gamma(s));\;b(e_1)=b(e_2)=b(e_3)=v}
\times\nonumber\\
&&\times
\mbox{Tr}(\tau_j (A(s(e_I)))^{-1}[A(s(e_I)),\hat{V}(v)^{1/2}]) \;\;
\mbox{Tr}(\tau_k (A(s(e_K)))^{-1}[A(s(e_K)),\hat{V}(v)^{1/2}]) \;\;
\times\nonumber\\
&&\times
\mbox{Tr}(\tau_l (A(s(e_L)))^{-1}[A(s(e_L)),\hat{V}(v)^{1/2}]) 
\ea
We have displayed the action for vertices of gravitational valence at 
least three. For gravitationally bi-valent vertices the action looks 
similar and is non-vanishing \cite{7.1}. These details will not be 
important for what follows.

Here $s(e)$ denotes an infinitesimal beginning segment of an edge $e$ and 
$\hat{V}(v)$ denotes the volume operator for an infinitesimal region $R(v)$
containing $v$. By the methods of \cite{7.1} it is easy to see that 
(\ref{7.18}) is a positive, essentially self-adjoint, gauge invariant and 
spatially diffeomorphism invariant operator on ${\cal H}_{Kin}$. Moreover,
the choice of the segments and regions $s(e),R(v)$ respectively is 
completely irrelevant as long as they contain $b(e),v$ respectively.

To solve the $\MCW$ (\ref{7.18}) is now surprisingly simple because 
the spectrum of the $\MCOW$ (\ref{7.18}) is {\it pure point}. Moreover, we 
can determine it sufficiently explicitly in order to actually derive the 
physical Hilbert space. To see this, notice that ${\cal H}^{YM}_{Kin}$ can
be decomposed as 
\be \label{7.19}
{\cal H}^{YM}_{Kin}=\overline{\oplus_{\gamma} {\cal 
H}^{YM}_{Kin;\gamma,\{\pi\}}}
\ee
where ${\cal H}^{YM}_{Kin;\gamma,\pi}$ is the finite 
linear 
span of charge network states $c$ with 
$\gamma(c)=\gamma,\pi(c)=\pi$ and the overline 
denotes closure. Now each ${\cal H}^{YM}_{Kin;\gamma,\pi}$ can be further 
decomposed as 
\be \label{7.20}
{\cal H}^{YM}_{Kin;\gamma,\pi}=\oplus_{\Pi} 
{\cal H}^{YM}_{Kin;\gamma,\pi,\Pi}
\ee
where $\Pi=\{\Pi_v\}_{v\in V(\gamma)}$ is a collection of 
equivalence classes of irreducible 
representations of $G$, one for each vertex of $v$, with the following 
meaning: From (\ref{7.5}) one easily verifies that the colour network 
state $T_c,\;\gamma(c)=\gamma,\pi(c)=\pi$ transforms 
under local $G-$gauge transformations with support at $v\in V(\gamma)$
in the representation 
\be \label{7.21}
[\otimes_{e\in E(\gamma);\;b(e)=v} \pi_e]\otimes
[\otimes_{e\in E(\gamma);\;f(e)=v} \pi^c_e]
\ee
where $\pi^c$ denotes the representation contragredient to $\pi$.
Since $G$ is compact, every representation is completely reducible and 
(\ref{7.21}) can be decomposed into mutually orthogonal but not 
necessarily inequivalent irreducible representations. Let the 
orthogonal projector on the representation space consisting of mutually 
orthogonal representations equivalent to the equivalence class $\Pi_v$
be denoted by 
\be \label{7.22}
i^{\Pi_v}_{[\otimes_{b(e)=v}\pi_e]\otimes[\otimes_{f(e)=v}\pi^c_e]}
\ee
also called an intertwiner. Notice that (\ref{7.22}) vanishes for all
$\Pi_v$ except for finitely many. We 
have the completeness relation 
\be \label{7.23}
\sum_{\Pi}\; 
i^{\Pi}_{[\otimes_{b(e)=v}\pi_e]\otimes[\otimes_{f(e)=v}\pi^c_e]}
=\mbox{id}_{[\otimes_{b(e)=v}\pi_e]\otimes[\otimes_{f(e)=v}\pi^c_e]}
\ee
Using (\ref{7.23}) for every $v\in V(\gamma)$ we can 
decompose each colour network state and arrive at (\ref{7.20}).
Finally we define 
\be \label{7.24}
{\cal H}^{YM}_{Kin;\gamma,\Pi}=\overline{\oplus_\pi
{\cal H}^{YM}_{Kin;\gamma,\Pi,\pi}}
\ee
and thus have the identity
\be \label{7.25}
{\cal H}^{YM}_{Kin}=\overline{\oplus_{\gamma,\Pi}
{\cal H}^{YM}_{Kin;\gamma,\Pi}}
\ee
The point of the decomposition (\ref{7.25}) is that the operator 
$\Delta_{\gamma,v}:=\delta^{JK} X^v_J X^v_K$ which appears in 
(\ref{7.18}) is diagonal on 
${\cal H}^{YM}_{Kin;\gamma,\Pi}$ with eigenvalue $-\lambda_{\Pi_v}\le 0$ 
given by the eigenvalue of the Laplace operator on $G$ in any
representation equivalent to $\Pi_v$. This also demonstrates that the sum 
in (\ref{7.25}) is indeed orthogonal. 
 
We can now proceed similarly with ${\cal H}^{GR}_{Kin}$ and decompose 
it as 
\be \label{7.26}
{\cal H}^{GR}_{Kin}=\overline{\oplus_{\gamma,\lambda} 
{\cal H}^{GR}_{Kin;\gamma,\lambda}}
\ee
where $-\lambda=\{-\lambda_v\}_{v\in V(\gamma)}\le 0$ are the eigenvalues 
of $(\widehat{e'(v)})^2$. That this operator is diagonizable in this 
fashion follows from a similar property for the volume operator itself.
The direct integral decomposition of ${\cal H}_{Kin}$ with respect 
to the $\MCOW$ is therefore simply 
\be \label{7.27}
{\cal H}_{Kin}=\overline{\oplus_{(\gamma,\Pi),(\gamma',\lambda)} 
{\cal H}^{YM}_{Kin;\gamma,\Pi}\otimes
{\cal H}^{GR}_{Kin;\gamma',\lambda}}
\ee
Since the spectrum of $\MCO$ is pure point we just need to identify the 
zero eigenspace in (\ref{7.27}) as the physical Hilbert space which, as a 
subspace of ${\cal H}_{Kin}$, carries the kinematical inner product 
as the physical inner product. To 
that end we notice that the eigenvalue on 
${\cal H}^{YM}_{Kin;\gamma,\Pi}\otimes{\cal H}^{GR}_{Kin;\gamma,\lambda}$
is given by 
\be \label{7.28}
\alpha^2(\frac{4}{\ell_p})^6\sum_{v\in V(\gamma)\cap 
V(\gamma')}\;\lambda_{\Pi_v} \lambda_v
\ee
If the $\lambda_v>0$ are not vanishing, the only way to make 
(\ref{7.28}) vanish for all $\gamma$ is to require $\lambda_{\Pi_v}=0$ for 
all
$v\in V(\gamma)$. Hence $\Pi_v=\mbox{Triv}$ must be the equivalence 
class of the trivial representation in that case. \\
\\
If $\lambda_v=0$
then $\Pi_v$ is arbitrary, which is different from what the ordinary
Gauss constraint would select. We actually do not know explicitly 
the space of states with $\lambda_v=0$ but it contains at least states 
with at least one vertex which is not at least trivalent with respect to 
the gravitational spin network structure. These zero eigenvalues 
are related to zero volume eigenstates because $e'(B)$ is related 
to the volume of the region $B$ and we know by now that there
are many such states even for higher valent vertices \cite{B}. 
To see that this does not pose any problem, notice that a zero volume 
vertex physically corresponds to a region which actually does not exist.
In other words, whether the Gauss constraint holds there or not is 
unaccessable by any observer\footnote{One could object that 
the zero volume region has a closed two -- boundary for which we could 
construct the electric flux operator. However, notice that in non -- 
Abelean gauge theories the electric flux operator for a closed two 
surface 1. does not measure the charge contained in the region bounded 
by the surface because it cannot be obtained by the (non -- Abelean)
Stokes theorem from the Gauss constraint and 2. it is not a Dirac 
observable even with respect to the Gauss constraint, that is, it is not 
gauge 
invariant and therefore does not correspond to a physical (measurable)
quantity. The area operator {\it is} gauge inavariant but does not measure
the charge neither, in fact, the charge is associated with vertices while 
the area is associated with edges. For Abelean gauge theories the 
objection is valid but here we are interested in a gravity coupled 
situation and then we have to consider also the spatial diffeomorphism 
and Hamiltonian constraints respectively for which the flux is not a 
Dirac observable again.}. Likewise, the results of \cite{B} 
demonstrate that the number of zero volume eigenvalues is neglible in a 
semiclassical sense because they correspond to spin configurations 
whose number compared to the number of all spin configurations at an 
$n-$valent vertex with maximal spin $j$ decreases roughly as $j^{-(n-1)}$ 
in the semiclassical limit of large quantum numbers $j$. Hence, 
semiclassical states are linear combinations of spin network states 
with almost always $\lambda_v>0$ as it should be since classical General 
Relativity is about non -- degenerate metrics. In what follows we will
simply drop that unphysical subsector of zero volume states for notational
simplicity.\\
\\
Then, denoting 
\be \label{7.29}
{\cal H}^{YM}_{Inv}:=\overline{\oplus_{\gamma} 
{\cal H}^{YM}_{Kin;\gamma,{\rm Triv}}}
\ee
the physical Hilbert space is given by the subspace
\be \label{7.30}
{\cal H}^{YM}_{Phys}={\cal H}^{YM}_{Inv}\otimes 
{\cal H}^{GR}_{Kin}\subset {\cal H}_{Kin}
\ee
consisting of the closed linear span of states $T_c\otimes T_s$ where 
$T_c$ is a gauge invariant charge network state constructed by using 
the intertwiner for the trivial representation. This is precisely the 
same phyiscal Hilbert space as selected by the Gauss constraint 
(\ref{7.8}) itself. 

\subsection{Einstein Gauss Constraint}
\label{s7.2}

We can proceed completely similarly with respect to the gravitational 
Gauss constraint for which the $\MCW$ is given by 
\be \label{7.31}
\MC:=\int_\sigma \;d^3x\; 
\frac{G_j G_k \delta^{jk}}{\sqrt{\det(q)}}
\ee
Its quantization proceeds entirely analogous to that of the 
Einstein -- Yang -- Mills Gauss constraint, just that it acts trivially
on the Yang -- Mills sector so that we will drop ${\cal H}^{YM}_{Kin}$
for the remainder of this section. Its action on spin network functions is 
given by
\be \label{7.32}
\MCO T_s =
\ell_p^4(\frac{4}{\ell_p})^6\sum_{v\in V(\gamma(s))}\;
\delta^{jk} X_j(v)
\widehat{e'(v)}^2 X_k(v) \;\;T_s
\ee
Now one could imagine that an ordering issue arises: In the ordering 
(\ref{7.32}) the operator is manifestly positive and essentially 
self-adjoint. However, in order to conclude similarly as in the previous 
section we need the ordering in which $\delta^{jk} X_j(v) X_k(v)$ stands 
to the outmost right. In the Yang -- Mills case there was no 
issue because the gravitational and Yang -- Mills degrees of 
freedom commute. Fortunately also here there is no problem because 
$X_j(v)$ is the generator of $SU(2)$ gauge transformations at $v$ and 
the operator $\widehat{e'(v)}$ is manifestly gauge invariant.
Hence the operators $\delta^{jk} X_j(v) X_k(v)$ and 
$\widehat{e'(v)}^2$ commute and can be diagonalized simultaneously.
 Hence we
can perform exactly the same steps as in section \ref{s7.1} and conclude 
that the physical Hilbert space is given by
\be \label{7.33}
{\cal H}_{Phys}={\cal H}_{Inv}\subset {\cal H}_{Kin}
\ee

\section{Infinite Number of Non-Abelean First Class Constraints 
Non -- Polynomial in the Momenta with Structure Functions}
\label{s8}

Euclidean 2+1 gravity can be formulated in complete analogy to Lorentzian
3+1 gravity (see e.g the fifth reference in \cite{7.1}). In this form 
the constraints are as difficult to solve as for the 3+1 theory which we 
reserve for future work \cite{8.1}. In particular, they are non-polynomial 
in the momenta and only close with structure functions rather than 
structure constants. However, there is a classically equivalent way 
(when the spatial metric is not degenerate) to write the constraints  
linear in the momenta. Namely Euclidean 2+1 gravity can be written as 
an $SU(2)-$gauge theory over a two dimensional Riemann surface 
$\sigma$ subject 
to a Gauss constraint and the curvature constraints
\be \label{8.1}
C^j=\frac{1}{2}\epsilon^{ab} F_{ab}^j 
\ee
where $F_{ab}^j$ is the curvature of $A_a^j$ and $\epsilon^{ab}$ is the 
density one valued skew tensor in two dimensions. It follows that 
(\ref{8.1}) is a scalar density of weight one. Accordingly we define the
corresponding $\MCW$ as 
\be \label{8.2}
\MC_E:=\frac{1}{2} \int_\sigma d^2x \frac{F^j 
F^k\delta_{jk}}{\sqrt{\det(q)}}
\ee
where $\det(q) q^{ab}=E^a_j E^b_k\delta^{jk}$ defines the spatial metric. 
Of course the pair $(A_a^j,E^a_j)$ is canonically conjugate up to the 
gravitational coupling constant $\kappa$. The significance of the label 
$E$ will be explained momentarily.

To quantize (\ref{8.2})
we will work directly on the space of gauge invariant states 
${\cal H}_{Inv}$ given by the closed linear span of gauge invariant spin 
network functions derived in the previous section. We proceed as in the 
fifth reference of \cite{7.1} and define the ``degeneracy vector'' 
\be \label{8.3}
E^j=\frac{1}{2} \epsilon_{ab}\epsilon^{jkl} E^a_k E^b_l 
\ee
called this way because $\det(q)=\delta_{jk} E^j E^k$. Using this we can 
establish an interesting identity:
Notice that 
the combinations (we use the identity $F_{ab}^j=\epsilon_{ab} F^j$)
\ba \label{8.3a}
C_a &:=& C_j \epsilon_{ab} E^b_j=F_{ab}^j E^b_j
\nonumber\\
C &:=& \frac{C_j E^j}{\sqrt{\det(q)}}
=\frac{1}{2} \frac{C_j \epsilon_{ab} E^a_k 
E^b_l\epsilon^{jkl}}{\sqrt{\det(q)}}
= \frac{1}{2}\frac{F_{ab}^j E^a_k E^b_l\epsilon^{jkl}}{\sqrt{\det(q)}}
\ea
look excacly like the spatial diffeomorphism and (Euclidean) 
Hamiltonian constraint of the 3+1 theory. Now one immediately verifies 
that
\be \label{8.3b}
\MC_E=\frac{q^{ab} C_a C_b+C^2}{\sqrt{\det(q)}}
\ee
Hence we arrive at the following crucial observation:\\
\\
{\bf The Master Constraint (\ref{8.2}) for the curvature constraints 
(\ref{8.1}) 
coincides with the extended $\MCW$ \cite{7.0} 
for the spatial Diffeomorphism and Hamiltonian constraints}.
\\
Hence 2+1 gravity not only tests the extended $\MCW$ idea but also can be 
considered as an example with a non -- Abelean constraint algebra with 
structure functions.

Similar to the 
previous section we define for a two dimensional region $R$ the following 
smeared quantities
\ba \label{8.4}
F_j(R) &=& \int_R d^2x F_j
\nonumber\\
E^j(R) &=& \int_R d^2x E^j
\nonumber\\
V(R) &=& \int_R d^2x \sqrt{\det(q)}
\ea
and write the classical $\MCW$ as the Riemann sum limit for a family 
of partitions $\cal P$, that is
\be \label{8.5}
\MC_E=\lim_{{\cal P}\to \sigma}\frac{1}{2}\sum_{R\in {\cal P}}
\frac{(F_j(R))^2 (E^k(R))^2}{(V(R))^3}
\ee
Using the classical identity for $x\in R$  
\be \label{8.6}
E^j(x)=\frac{1}{\kappa^2} \epsilon^{ab}\epsilon_{jkl}
\{A_a^k(x),V(R)\}\{A_b^l(x),V(R)\}
\ee
we have 
\be \label{8.7}
\frac{E^j(R)}{V(R)^{3/2}}=(\frac{4}{\kappa})^2\epsilon_{jkl}
\int_R \{A^k,V(R)^{1/4}\}\wedge\{A^l,V(R)^{1/4}\}
=:(\frac{4}{\kappa})^2 E'_j(R)
\ee
Hence the $\MCW$ becomes 
\be \label{8.8}
\MC_E=\lim_{{\cal P}\to \sigma}\frac{1}{2}(\frac{4}{\kappa})^4
\sum_{R\in {\cal P}} (F_j(R))^2 (E'_k(R))^2
\ee
Again we specialize to a simplicial decomposition ${\cal P}$. We single 
out a corner $v(R)$ for each triangle $R$ and denote by $e_I(R)$ the two 
edges of $\partial R$ starting at $v(R)$. Then 
\ba \label{8.9}
\MC_E &=& \lim_{{\cal P}\to \sigma} 2(\frac{4}{\kappa})^4
\sum_{R\in {\cal P}} (\tilde{F}_j(R))^2 (\tilde{E}'_k(R))^2
\\
\tilde{F}_j(R) &=& \mbox{Tr}(\tau_j A(\partial R))
\nonumber\\
\tilde{E}'_j(R) &=& \epsilon^{jkl}\sum_{K,L=1,2} 
\epsilon^{IJ}
\times\nonumber\\
&&\times
\mbox{Tr}(\tau_k A(e_K(R))\{A(e_K(R))^{-1},V(R)^{1/4}\})\;\;
\mbox{Tr}(\tau_l A(e_L(R))\{A(e_L(R))^{-1},V(R)^{1/4}\})
\nonumber
\ea
Now (\ref{8.9}) is written in terms of holonomies and the volume operator 
both of which admit well-defined quantizations on ${\cal H}_{Kin}$ 
(see the fifth reference of \cite{7.1} for the 2+1 volume operator). 

There is a difference in formulating the $\MCOW$ $\MCO$ for the Hamiltonian 
constraint and the extended $\MCOW$ $\MCO_E$ for the combined spatial 
diffeomorphism 
and Hamiltonian constraint: As shown in \cite{7.0} the former can only be 
defined, in its fundamental form, on the spatially diffeomorphism 
invariant Hilbert space ${\cal H}_{Diff}$ \cite{8.2}
of solutions to the spatial
diffeomorphism constraint while the second of course must be defined 
on the kinematical Hilbert space ${\cal H}_{Kin}$ since it solves both 
types of constraints in one step. However, since $\MCO_E$ is 
nevertheless a spatially diffeomorphism invariant operator, it can be 
defined on ${\cal H}_{Kin}$ only if it does not change the graph of a spin 
network state on which it acts as was shown in detail \cite{8.2}. 
We must keep this in mind when quantizing (\ref{8.9}).

Our heuristic ansatz will be to define the following quadratic form 
in which we order the various operator factors judiciously
(if it exists)
\ba \label{8.10}
&& Q_{\MC_E}(T_s,T_{s'}) 
= \lim_{{\cal P}\to 
\sigma} 2(\frac{4}{\ell_p^2})^4
\sum_{R\in {\cal P}} \delta^{jk} \delta^{mn}
\sum_{s_1}
\times\\
&&\times
<T_s,\widehat{\tilde{F}_j(R)} \widehat{\tilde{E}'_m(R)} T_{s_1}>_{Kin}
\overline{<T_{s'},\widehat{\tilde{F}_k(R)} \widehat{\tilde{E}'_n(R)} 
T_{s_1}>_{Kin}}
\nonumber\\
&=& 2(\frac{4}{\ell_p^2})^4 \sum_{s_1}
\lim_{{\cal P}\to \sigma} 
\sum_{R\in {\cal P}} \delta^{jk} \delta^{mn}
\times\nonumber\\
&&\times
<T_s,\widehat{\tilde{F}_j(R)} \widehat{\tilde{E}'_m(R)} T_{s_1}>_{Kin}\;\;
\overline{<T_{s'},\widehat{\tilde{F}_k(R)} \widehat{\tilde{E}'_n(R)} 
T_{s_1}>_{Kin}}
\mbox{ where}
\nonumber\\
&& \widehat{\tilde{E}'_j(R)} T_{s_1} = \epsilon^{jkl}\sum_{K,L=1,2} 
\epsilon^{KL}
\times\nonumber\\
&&\times
\mbox{Tr}(\tau_k A(e_K(R))[A(e_K(R))^{-1},\hat{V}(R)^{1/4}])\;\;
\mbox{Tr}(\tau_l A(e_L(R))[A(e_L(R))^{-1},\hat{V}(R)^{1/4}])\;\; T_{s_1}
\nonumber
\ea
and $\widehat{\tilde{F}_j(R)}$ acts by multiplication. Notice that the sum 
over $s_1$ is an uncountably infinite sum but we will see soon that for 
given $s,s'$ only a finite number of terms contribute so that the 
interchange of the limit with the sum in the second step is indeed 
justified. One may wonder why 
we have introduced this insertion of unity at all. This is motivated by
the theory in 3+1 dimensions \cite{7.0,8.1} where this insertion of unity
is mandatory when working at the level of ${\cal H}_{Diff}$ and where
$T_s,T_{s'}$ are replaced by diffeomorphism invariant distributions.

It remains to take the limit ${\cal P}\to \sigma$. First of all,
as the regions $R$ shrink to points, every term in the sum over the $R$
at given $s_1$
vanishes anyway if $\gamma(s)\not=\gamma(s')$  
due to the orthogonality of the spin network functions so that
both graphs must be subgraphs of $\gamma(s_1)\cap \partial R$, so they
must equal each other for sufficiently small $R$. Thus:\\
\\
{\bf The $\MCO_E$ regularized this way
in terms of a quadratic form is automatically not graph changing}.\\
\\
Next, no matter how the limit ${\cal P}\to
\sigma$ is performed, given $s$, for sufficiently fine $\cal P$ the last 
line in (\ref{8.10}) vanishes unless $R$ intersects $\gamma(s)$. 
Now in contrast to the situation in 3+1 dimensions, if $R$ intersects
$\gamma(s)$ but does not contain a vertex of $\gamma(s)$, the last line in 
(\ref{8.10}) is not automatically vanishing which leads to a divergence 
when the limit ${\cal P}\to \sigma$ is not carefully performed.  
Following the proposal in the fifth 
reference of \cite{7.1} we adapt the limit ${\cal P}\to\sigma$ for 
the matrix element (\ref{8.10}) to the states $T_{s_1}$ in the 
following sense:\\ 
1.\\
The triangles $R$ are chosen to saturate each vertex $v$
of $\gamma(s_1)$, that is, for each adjacent pair of edges $e,e'$ there 
is precisely one triangle $R$ with $v(R)=v$ and $e_1(R)\subset 
e,e_2(R)\subset e'$.\\
2.\\
Away from the vertices, the triangles $R$ intersect $\gamma(s)$ only in 
such a way that the vertices of the graph $\partial R\cap \gamma(S)$
which are not vertices of $\gamma(s)$ 
are co-linear, that is, the tangents of the respective intersecting edges 
are linearly dependent there.\\
\\
This results in a different limit ${\cal P}\to \sigma$ for each 
$\gamma(s_1)$
which is justified by the fact that classically the Riemann sum converges 
to the same limit no matter which sequence is chosen. The second 
requirement now makes sure that the only contributions come from the 
triangles with $v(R)\in V(\gamma(s))$, hence (\ref{8.10}) simplifies to
\ba \label{8.11}
Q_{\MC_E}(T_s,T_{s'}) &=& \delta_{\gamma(s),\gamma(s')} \sum_{s_1}
\lim_{{\cal P}\to \sigma} 
2(\frac{4}{\ell_p^2})^4 \sum_{v\in V(\gamma(s))}\;\;\;
\sum_{R\in {\cal P};\;v(R)=v} \delta^{jk} \delta^{mn}
\times\nonumber\\
&&\times
<T_s,\widehat{\tilde{F}_j(R)} \widehat{\tilde{E}'_m(R)} T_{s_1}>_{Kin}\;\;
\overline{<T_{s'},\widehat{\tilde{F}_k(R)} \widehat{\tilde{E}'_n(R)} 
T_{s_1}>_{Kin}}
\ea
Now a miracle happens: The limit involved in (\ref{8.11}) is already 
trivial because, as the triangles $R(v)$ shrink, this shrinking 
can be absorbed by an (analytical) diffeomorphism of $\sigma$ which 
preserves $\gamma(s)=\gamma(s')$. However, since the diffeomorphism group 
is 
represented unitarily on ${\cal H}_{Kin}$ \cite{8.2}, the number 
on the right hand side of (\ref{8.11}) actually no longer depends on the 
``size and shape'' of the triangles $R$ but only on their diffeomorphism 
invariant characteristics (this is easiest to see by performing the sum 
over $s_1$ thus resulting in a unit operator). 
This reduces the number of regularization ambiguities
tremendously from uncountably infinite to countably infinite (namely
the $C^n$ class of the intersections at the two vertices of $\partial R 
\cap \gamma(s)$ different from $v(R)$). 

Hence, we are left with making that choice. The most natural choice is the 
one made in \cite{7.1} namely such that the the edge $\partial 
R-\gamma(s_1)$
intersects $\gamma(s_1)$ transversally at both end points (corresponding 
to 
a $C^0$ intersection). However, it is easy to see that the limit in 
(\ref{8.11}) is not satisfactory unless $\partial R\subset \gamma(s)$.
The reason is that up to the commutator between  
$\widehat{\tilde{F}_k(R)},\widehat{\tilde{E}'_n(R)}$ (which is
of higher order in $\hbar$) after 
summation over 
$s_1$ the matrix element is proportional to a sum of matrix elements 
between spin network states over $\gamma(s)$ of the operator
$$
(\widehat{\tilde{F}_j(R)})^\dagger
\widehat{\tilde{F}_j(R)}=3-\chi_1(\hat{A}(\partial R))
$$
where $\chi_j$ is the character of the irreducible representation of 
$SU(2)$ with weight $j$. Since $\gamma(s)=\gamma(s')$ and since 
$\widehat{\tilde{E}'_j(R)}$ only acts on the intertwiners of a spin-network
state but not the spins, it follows that the piece 
$\chi_1(\hat{A}(\partial R))$
drops out of (\ref{8.11}) for sufficiently small $\partial R$. Hence the 
matrix element would not contain any information about the curvature 
$F_j$. This happens because we are working at the kinematical level and 
not at the spatially diffeomorphism invariant level: If $T_{s}$ would 
be diffeomorphism invariant and depend only on the diffeomorphism class
of $s$ \cite{8.1} then there would be a non-vanishing contribution no matter 
how small $\partial R$ because we could for instance choose $s$ to be in 
the class of the spin network state $\chi_1(A(\partial R)) T_{s'}$.  
 
It follows that at the level of ${\cal H}_{Kin}$ we must proceed 
differently than for ${\cal H}_{Diff}$ in order to obtain a satisfactory 
operator. The necessary idea was outlined already in \cite{7.0,8.3}:
\begin{Definition} \label{def8.1} ~~~\\
i)\\
Let $\gamma$ be a graph, $v\in V(\gamma)$ a vertex and $e,e'\in E(\gamma)$
two different edges starting at $v$ (reverse orientation if necessary). A 
loop $\alpha_{\gamma,v,e,e'}$ within $\gamma$ starting at $v$ along 
$e$ and ending at $v$ along $(e')^{-1}$ is said to be minimal if there
is no other loop within $\gamma$ with the same properties and less edges 
of $\gamma$ traversed.\\
ii)\\
Given the data $\gamma,v,e,e'$ we denote by $L(\gamma,v,e,e')$ the set
of minimal loops compatible with those data.
\end{Definition}
Notice that the notion of a minimal loop does not refer to a background
structure such as a metric. The set $L(\gamma,v,e,e')$ is never empty
for the closed graphs that we are considering here and it is always 
finite.

The idea is now that semiclassical states must necessarily have very 
complex graphs in order that, e.g. the volume operator for every 
macroscopic 
region has non-zero expection values. Moreover, they will involve 
very high spins because high spin means large volume so that the 
correspondence principle is satisfied. For such 
complex graphs the 
limiting loops $\partial R$ are of the same order of magnitude as the 
minimal loops associated with it. Hence on semiclassical states it makes 
sense to replace the limit by those minimal loops. This is justified in 
the sense that we are quantizing an operator corresponding to a classical
quantity which we can hope to be approximated well for semiclassical 
states only. Since then there is no consistency check for the details at 
the microscopic level, our procedure is justified in this semiclassical 
sense. 
Of course, the argument is not entirely satisfactory and can at best
result in an effective description $\MCO_{E,eff}$ of the fundamental 
operator $\MCO_E$ which, however, must then be defined on ${\cal 
H}_{Diff}$
rather than ${\cal H}_{Kin}$ in order to obtain a non-trivial result. 
Hence, at the level of ${\cal H}_{Kin}$ this is the best we can do.
The more fundamental programme will be carried out in \cite{8.1} directly 
for the 3+1 case.

Summarizing, the close to final matrix element is defined by
\ba \label{8.12}
&&Q_{\MC_{E,eff}}(T_s,T_{s'}) = \delta_{\gamma(s),\gamma(s')}
\times\\ && \times
\sum_{s_1}
2(\frac{4}{\ell_p^2})^4 \sum_{v\in V(\gamma(s))}
\delta^{jk} \delta^{mn}
\sum_{e_1,e_2\in E(\gamma(s));v=e_1\cap e_2}\;\;
\sum_{\alpha\in L(\gamma(s),v,e_1,e_2)}
\frac{1}{|L(\gamma(s),v,e_1,e_2)|}
\times\nonumber\\ &&\times
<T_s,\widehat{\tilde{F}_j(R(\alpha))} \widehat{\tilde{E}'_m(v,e_1,e_2)} 
T_{s_1}>_{Kin}\;\;
\overline{<T_s',\widehat{\tilde{F}_k(R(\alpha))} 
\widehat{\tilde{E}'_n(v,e_1,e_2)} T_{s_1}>_{Kin}}
\nonumber\\
&&\widehat{\tilde{E}'_j(v,e,e')} T_{s_1} = \epsilon^{jkl}\sum_{K,L=1,2} 
\epsilon^{KL}
\times\nonumber\\ &&\times
\mbox{Tr}(\tau_k A(s(e_K))[A(s(e_K))^{-1},\hat{V}(v)^{1/4}])\;\;
\mbox{Tr}(\tau_l A(s(e_L))[A(s(e_L))^{-1},\hat{V}(v)^{1/4}])\;\; T_s
\nonumber
\ea
Here we have averaged over the number of minimal loops with data 
$\gamma(s),v,e,e'$ in order not to overcount as compared to (\ref{8.11})
and $R(\alpha)$ is the unique, interior, {\bf contractable} region 
enclosed by $\alpha$, i.e. $\partial R(\alpha)=\alpha$. As before 
$s(e)$ denotes an infinitesimal beginning segment of an edge $e$ and 
$\hat{V}(v)$ the volume operator for an infinitesimal region containing 
$v$. 
Notice that (\ref{8.12}) is a drastic modification of (\ref{8.11}) 
unless the graph $\gamma(s)$ is very complex, hence we have put the 
subscript ``effective''. 

Formula (\ref{8.12}) is not yet quite what we want because we have to 
impose the constraint $\delta_{\gamma(s),\gamma(s')}$ explicitly. We can 
avoid that by inserting a projection operator $\hat{P}_\alpha$ which 
projects onto the closed subspace of spin network states which have 
spin higher than $1/2$ for all edges of $\alpha$. This again modifies 
$\MCO_E$ for spin network states involving low spins but does not change 
its semiclassical behaviour for the reason mentioned above. 
Hence the final expression for the effective $\MCOW$ is 
\ba \label{8.12a}
&&Q_{\MC_{E,eff}}(T_s,T_{s'}) = 
\sum_{s_1}
2(\frac{4}{\ell_p^2})^4 \sum_{v\in V(\gamma(s))}
\delta^{jk} \delta^{mn}
\times\\ &&\times
\sum_{e_1,e_2\in E(\gamma(s));v=e_1\cap e_2} \;\;\sum_{\alpha\in 
L(\gamma(s),v,e_1,e_2)}
\frac{1}{|L(\gamma(s),v,e_1,e_2)|}
\times\nonumber\\ &&\times
<T_s,\widehat{\tilde{F}_j(R(\alpha))} \widehat{\tilde{E}'_m(v,e_1,e_2)} 
\hat{P}_\alpha T_{s_1}>_{Kin}\;\;
\overline{<T_s',\widehat{\tilde{F}_k(R(\alpha))} 
\widehat{\tilde{E}'_n(v,e_1,e_2)} 
\hat{P}_\alpha T_{s_1}>_{Kin}}
\nonumber
\ea
Due to the projections the state 
$$
\widehat{\tilde{F}_j(R(\alpha))} \widehat{\tilde{E}'_m(v,e_1,e_2)} 
\hat{P}_\alpha T_{s_1}
$$
is a linear combination of spin network states over the graph 
$\gamma(s_1)\cup \alpha=\gamma(s_1)$, hence we must have 
$\gamma(s)=\gamma(s_1)=\gamma(s')$ in order that the matrix element for 
$s_1,\alpha$
does not vanish. Thus the condition $\gamma(s)=\gamma(s')$ is implicit.

Expression (\ref{8.12a}) defines a positive quadratic form. To see the 
positivity we take a generic linear combination of 
spin network functions 
$$
f=\sum_{p=1}^M \sum_{r=1}^{N_p} z_p^r T_{s_p^r}
$$
where $z_p^r$ are complex numbers and we have adopted a labelling such 
that $\gamma(s_p^r)=\gamma_p$ for all $r=1,..,N_p$ and the graphs 
$\gamma_p$ are mutually different. Then we simply compute, using the fact
that the matrix elements vanish between spin network states over different 
graphs 
\ba \label{8.12b}
&& Q_{\MC_{E,eff}}(f,f)
= \sum_{p=1}^N \sum_{r,s=1}^{N_p} \bar{z}_p^r z_p^l
\times\\ &&\times
\sum_{s_1}
2(\frac{4}{\ell_p^2})^4 \sum_{v\in V(\gamma_p)}
\delta^{jk} \delta^{mn}
\sum_{e_1,e_2\in E(\gamma_p);v=e_1\cap e_2} \;\;\sum_{\alpha\in 
L(\gamma_p,v,e_1,e_2)}
\frac{1}{|L(\gamma_p,v,e_1,e_2)|}
\times\nonumber\\ &&\times
<T_{s_p^r},\widehat{\tilde{F}_j(R(\alpha))} 
\widehat{\tilde{E}'_m(v,e_1,e_2)} 
\hat{P}_\alpha T_{s_1}>_{Kin}\;\;
\overline{<T_{s_p^s},\widehat{\tilde{F}_k(R(\alpha))} 
\widehat{\tilde{E}'_n(v,e_1,e_2)} 
\hat{P}_\alpha T_{s_1}>_{Kin}}
\nonumber\\
&=& \sum_{p=1}^N 
\sum_{s_1}
2(\frac{4}{\ell_p^2})^4 \sum_{v\in V(\gamma_p)}
\sum_{j,m}
\sum_{e_1,e_2\in E(\gamma_p);v=e_1\cap e_2} \;\;\sum_{\alpha\in 
L(\gamma_p,v,e_1,e_2)} \frac{1}{|L(\gamma_p,v,e_1,e_2)|}
\times\nonumber\\ &&\times
|\sum_{r=1}^{N_p} \bar{z}_p^r <T_{s_p^r},\widehat{\tilde{F}_j(R(\alpha))} 
\widehat{\tilde{E}'_m(v,e_1,e_2)} 
\hat{P}_\alpha T_{s_1}>_{Kin}|^2
\nonumber
\ea
which is manifestly non -- negative.

Not every 
positive quadratic form defines an operator but if it does (technically,
if it is closable), then the 
corresponding positive, self-adjoint operator is unique (i.e. there is no
choice in its self-adjoint extension). 
\begin{Theorem} \label{th8.1}  ~~~~~~\\
The quadratic from (\ref{8.12a}) defines a positive self -- adjoint 
operator.
\end{Theorem}
Proof of theorem \ref{th8.1}:\\
To see this we notice that the would be operator is given by
\be \label{8.13}
\MCO_{E,eff} T_s=\sum_{s'} Q_{\MC_{E,eff}}(T_{s'},T_s) T_{s'}
\ee
This defines a (densely defined) operator if and only if the right hand 
side of (\ref{8.13}) is normalizable, that is,
\be \label{8.14}
||\MCO_{E,eff} T_s||^2_{Kin}=\sum_{s'} |Q_{\MC_{E,eff}}(T_{s'},T_s)|^2
\ee
converges. Notice that the sum on the right hand side is over an 
uncountably infinite set, so (\ref{8.13}) looks dangerous. However, 
we know already that (\ref{8.14}) reduces to the 
countable sum
\be \label{8.15}
||\MCO_{E,eff} T_s||^2_{Kin}=\sum_{\gamma(s')=\gamma(s)} 
|Q_{\MC_{E,eff}}(T_{s'},T_s)|^2
\ee
Now the matrix elements (\ref{8.12a}) are finite for given $s,s'$ since 
we must have $\gamma(s_1)=\gamma(s)$
and the states 
$\widehat{\tilde{F}_j(R(\alpha))} \widehat{\tilde{E}'_m(v,e_1,e_2)} 
\hat{P}_\alpha T_{s_1}$
are a finite linear combination of spin network states of which at most 
one can coincide with either $T_s$ or $T_{s'}$. Since  
there are 
only $\sum_{v\in V(\gamma(s))} n(v)$ such terms involved in (\ref{8.12a})
where $n(v)$ denotes the valence of $v$, the assertion follows. 

Hence, to show convergence of 
(\ref{8.15}) it is sufficient to show that for given $s$ there are only a 
finite number of $s'$ for which $Q_{\MC_{E,eff}}(T_{s'},T_s)\not=0$.
Now we have seen that $Q_{\MC_{E,eff}}(T_{s'},T_s)\not=0$ implies that
in the sum over $s_1$ only a finite number of terms contribute which
must satisfy $s_1\in S(s)$ where $S(s)$ is a finite set of spin network 
labels. For each $s_1\in S(s)$ consider the finite number of spin network 
states with labels in $S(s_1)$ contained in the span of the states  
$$
\widehat{\tilde{F}_j(R(\alpha))} \widehat{\tilde{E}'_m(v,e_1,e_2)} 
\hat{P}_\alpha T_{s_1}
$$
as $v,e_1,e_2,\alpha\in L(\gamma(s),v,e_1,e_2)$ vary. Thus $s'$ must be in 
the finite set $\cup_{s_1\in S(s)}S(s_1)$.

This shows that (\ref{8.13}) defines a positive symmetric operator on the 
finite linear span of spin network states. However, positive symmetric
operators have a preferred self -- adjoint extension, the Friedrichs 
extension, which is given by the unique closure of the 
corresponding quadratic form $Q_{\MC_{E,eff}}$ with the same domain of 
definition.\\
$\Box$\\
\\
We now can proceed to solve $\MCO_{E,eff}$. 
Let $\Phi_{Kin}$ be the dense subset of 
${\cal H}_{Kin}$ given by the finite linear span of spin network 
functions. 
We begin by verifying that for any $f\in \Phi_{Kin}$ the distribution
\be \label{8.17}
\eta(f):=\bar{f}\delta[F],\;\; \delta[F]:=
\int_{{\cal M}_{flat}} d\nu_0(A_0)  \;\delta_{A_0}
\ee
solves the effective extended $\MCOW$. Here ${\cal M}_{flat}$ is the 
moduli 
space of flat connections on $\sigma$, $\delta_{A_0}$ is the 
$\delta-$distribution supported at $A_0$ and $\nu_0$ is the following 
measure on ${\cal M}_{flat}$: Any function on ${\cal M}_{flat}$ is
of the form $f(A_0)=f_n(A_0(\alpha_1),..,\alpha_n(A_0))$ where 
$\alpha_1,..,\alpha_n$ are any generators of the fundamental group
$\pi_1(\sigma)$. Then 
\be \label{8.18}
\nu_0(f)=\int_{{\cal M}_{flat}} d\nu_0(A_0) f(A_0)=\int_{SU(2)^n}\;\;
d\mu_H(h_1).. d\mu_H(h_n)\;\; f_n(h_1,..,h_n) 
\ee
The interested reader is referred to \cite{7.1} for more details.

Now we may check whether (\ref{8.17}) is a generalized solution of
$\MCO_{E,eff}=0$. We have for any $f'\in \Phi_{Kin}$
\ba\label{8.19}
\eta(f)[\MCO_{E,eff} f'] &=&
\int_{{\cal M}_{flat}}\;\; d\nu_0(A_0)\;\overline{f(A_0)} 
\;\;\delta_{A_0}[\MCO_{E,eff} f']
\nonumber\\
&=&
\int_{{\cal M}_{flat}}\;\; d\nu_0(A_0)\;\overline{f(A_0)} 
[\MCO_{E,eff} f'](A_0)
\ea
This will be zero if and only if it is zero for any spin network 
state $f'=T_s$, hence we compute 
\ba\label{8.19a}
&&\eta(f)[\MCO_{E,eff} T_s] = 
\sum_{s'} 
<f,T_{s'}>_{Flat} Q_{\MCO_{E,eff}}(T_{s'},T_s)
\nonumber\\
&=& 2(\frac{4}{\ell_p^2})^4 
\sum_{s',s_1} <f,T_{s'}>_{Flat} \sum_{v\in V(\gamma(s))}
\delta^{jk} \delta^{mn}
\times\nonumber\\&&\times
\sum_{e_1,e_2\in E(\gamma(s));v=e_1\cap e_2} \;\;\;\sum_{\alpha\in 
L(\gamma(s),v,e_1,e_2)}
\frac{1}{|L(\gamma(s),v,e_1,e_2)|}
\times\nonumber\\ &&\times
<T_{s'},\widehat{\tilde{F}_j(R(\alpha))} 
\widehat{\tilde{E}'_m(v,e_1,e_2)} 
\hat{P}_\alpha T_{s_1}>_{Kin}\;\;
\overline{<T_s,\widehat{\tilde{F}_k(R(\alpha))} 
\widehat{\tilde{E}'_n(v,e_1,e_2)} 
\hat{P}_\alpha T_{s_1}>_{Kin}}
\nonumber\\
&=& 2(\frac{4}{\ell_p^2})^4 
\sum_{s_1}  \sum_{v\in V(\gamma(s))}
\delta^{jk} \delta^{mn}
\sum_{e_1,e_2\in E(\gamma(s));v=e_1\cap e_2} \sum_{\alpha\in 
L(\gamma(s),v,e_1,e_2)}
\frac{1}{|L(\gamma(s),v,e_1,e_2)|}
\times\nonumber\\ &&\times
<f,\widehat{\tilde{F}_j(R(\alpha))} 
\widehat{\tilde{E}'_m(v,e_1,e_2)} 
\hat{P}_\alpha T_{s_1}>_{Flat} \;\;
\overline{<T_s,\widehat{\tilde{F}_k(R(\alpha))} 
\widehat{\tilde{E}'_n(v,e_1,e_2)} 
\hat{P}_\alpha T_{s_1}>_{Kin}}
\nonumber\\
&=& 0
\ea
In the last step we have used the completeness relation with respect to
$s'$ and the fact that
Tr$(\tau_j A_0(\alpha))=0$ for every $A_0\in {\cal M}_{flat}$. We could 
interchange the various summations because the non-vanishing terms reduce
the sums to finite ones. 

It follows that
\be \label{8.20}
\MCO_{E,eff}' \eta(f)=0
\ee
where the prime denotes the dual of the operator on the space 
$\Phi_{Kin}^\ast$ of linear functionals on $\Phi_{Kin}$ (without 
continuity requirement) defined in general by $[O' l](f)=l[O^\dagger f]$.
Hence we see that the point $\lambda=0$ lies at least in the continuous 
part of $\MCO_{E,eff}$ since the $\eta(f)$ are not normalizable 
with respect to $<.,.>_{Kin}$, provided that we can write 
$\eta(f)[f']=<\tilde{f}(0),\tilde{f}'(0)>_{{\cal H}^{c\oplus}_0}$ 
where $\tilde{f},\;\tilde{f}'$ are representatives of $f,f'$ corresponding 
to a direct integral decomposition of $\cal H$ subordinate to
$\MCO_{E,eff}$. One should now complete the analysis and 
compute the full spectrum of $\MCO_{E,eff}$ to see whether that is indeed 
the case. Assuming that to be true, for the purposes of this 
paper it is sufficient to note that whatever the complete ${\cal 
H}_{Phys}$ might be, it contains the closed linear span of the 
$\eta(f)$ as a subspace with the induced physical inner product
\be \label{8.21}
<\eta(f),\eta(f')>_{Phys}:=\nu_0(f\overline{f'})
\ee
which is of course well known in the literature \cite{8.4}. We leave a 
more complete analysis for future work and just remark that with the 
techniques of \cite{8.5} it is not difficult to show that the operator 
(\ref{8.13}), although we have performed rather drastic manipulations, 
indeed has the correct classical limit.

\section{Conclusions and Outlook}
\label{s9}

What we have learnt in this paper is that the Master Constraint Programme
is also able to deal with the case of interacting quantum field theories.
By this we mean that we can solve, e.g. the Gauss contraint of Non -- 
Abelean Yang Mills theory {\it when coupled to gravity}. Classical Non -- 
Abelean Yang -- Mills theory on a background spacetime is a self -- 
interacting field theory and in four dimensions nobody was able to show 
that the corresponding interacting quantum field theory exists (in the 
continuum; on the lattice there are no problems). The Gauss constraint 
of the theory is a quadratic polynomial in the fields and its square
id a fourth order polynomial. Therefore, from a QFT on curved background 
point of view the Master Constraint should be as UV singular as the 
square of the curvature that defines the Yang -- Mills action. Things 
become even worse because on top of that we have multiplied the square of 
the Gauss constraint by a factor which depends non -- polynomially on the 
the degrees of freedom of the gravitational field. 

Yet, we were able 
to quantize the integral of the resulting expression on the Hilbert space
which is used in Loop Quantum Gravity (LQG) without encountering UV
problems. The technical reason for why this happened is that the 
Master Constraint is spatially diffeomorphism invariant integral of a 
scalar density. As was shown in the sixth reference of \cite{7.1}, for 
such quantities, in a very precise sense, the UV regulator gets swallowed 
up by spatial 
diffeomorphism group: The theory does not depend on a background metric, 
thus all ``distances'' are gauge equivalent. 

Let us compare the situation 
with our previous paper \cite{IV}: There we were looking at free field 
theories on a Minkowski background and we could use the associated 
background dependent Fock representations. The square of the Maxwell
Gauss constraint is too singular and cannot be employed to define the 
Master Constraint, thus we had to use the flexibility of the Master 
Constraint Programme to regularize the Master Constraint by a background 
dependent kernel. In the present situation the singularity of the square
is expected, according to perturbation theory arguments, to be even worse.
However, fundamentally the Fock representation is not a valid 
representation for interacting quantum field theories. A possible
representation is the one that uses LQG techniques, however, that 
representation is only valid when we couple the gravitational field as 
otherwise e.g. the Yang -- Mills Hamiltonian is ill -- defined in that
representation. 

It is precisely this observation which has lead us to 
quantize the Master Constraint in this kind of LQG representation.
It turns out that the non -- polynomial, gravitational field dependent 
function mentioned above that enters the Master Constraint plays exactly 
the same role as for the Yang -- Mills Hamiltonian (constraint): It serves 
as a background independent UV regulator. Moreover, while in the free 
field case
the regulator could be chosen to be background dependent, here we are not 
allowed to do that, however, {\it background independent theories, through 
gravity, have the 
tendency to regulate themselves}. Thus the factor of 
$1/\sqrt{\det(q)}$ which typically enters the background independent 
Master Constraints becomes a quantum operator (rather than a $\Cl$ number 
valued expression) but otherwise plays the same role as the trace class 
operator $K$ in \cite{IV}: It removes the UV singularities of the square 
of the Gauss constraints.

We stress again that this does not mean at all that we have proved the 
existence of, say, QCD. This is because, while we are able to avoid 
certain singularities, now the burden is on us to show that the theory
can also successfully deal with the additional symmetries that have 
entered the stage by coupling matter to the gravitational field. This
is the spacetime diffeomorphism symmetry which finds its way into the 
canonical framework in the form of the spatial diffeomorphism and the 
Hamiltonian constraint. It is precisely for this reason that the Master
Constraint Programme was created. One now has to apply it to all 
symmetries of General Relativity, solve the full Master Constraint and 
establish that we have captured a quantum theory of General Relativity 
rather than a mathematically consistent but physically uninteresting
quantum theory of geometry and matter. This is what has to be done in the 
close future and finally the mathematical techniques are available 
in order to make progress. \\
\\
\\
\\
{\large Acknowledgements}\\
\\
BD thanks the German National Merit Foundation for financial support.
This research project was supported in part by a grant from 
NSERC of Canada to the Perimeter Institute for Theoretical Physics.


\begin{thebibliography}{99}

\parskip -5pt

\bibitem{7.0} T. Thiemann, ``The Phoenix Project: Master Constraint 
Programme for Loop Quantum Gravity'', gr-qc/0305080

\bibitem{I} B. Dittrich, T. Thiemann, ``Testing the Master Constraint 
Programme for Loop Quantum Gravity I. General Framework'', 
gr-qc/0411138

\bibitem{II} B. Dittrich, T. Thiemann, ``Testing the Master Constraint 
Programme for Loop Quantum Gravity II. Finite Dimensional Systems'',
gr-qc/0411139

\bibitem{III} B. Dittrich, T. Thiemann, ``Testing the Master Constraint 
Programme for Loop Quantum Gravity III. SL(2,R) Models'',
gr-qc/0411140

\bibitem{IV} B. Dittrich, T. Thiemann, ``Testing the Master Constraint 
Programme for Loop Quantum Gravity IV. Free Field Theories'',
gr-qc/0411141

\bibitem{1.1} 
C. Rovelli, ``Loop Quantum Gravity", Living Rev. Rel. {\bf 1} (1998) 1,
gr-qc/9710008\\
T. Thiemann,``Lectures on Loop Quantum Gravity'', Lecture Notes in 
Physics, {\bf 631} (2003) 41 -- 135, gr-qc/0210094\\
A. Ashtekar, J. Lewandowski, ``Background Independent Quantum Gravity:
A Status Report'', Class. Quant. Grav. {\bf 21} (2004) R53; 
[gr-qc/0404018]\\
L. Smolin, ``An Invitation to Loop Quantum Gravity'', hep-th/0408048

\bibitem{7.2} C. Rovelli, ``Quantum Gravity'', Cambridge University Press,
Cambridge, 2004

\bibitem{7.3} T. Thiemann, ``Modern Canonical Quantum General 
Relativity'', Cambridge University Press, Cambridge, 2005,
gr-qc/0110034

\bibitem{7.1} T. Thiemann, ``Anomaly-free Formulation of non-perturbative,
four-dimensional Lorentzian Quantum Gravity", Physics Letters {\bf B380}
(1996) 257-264, [gr-qc/9606088]\\
T. Thiemann, ``Quantum Spin Dynamics (QSD)",
Class. Quantum Grav. {\bf 15} (1998) 839-73, [gr-qc/9606089];
``II. The Kernel of the Wheeler-DeWitt Constraint Operator",
Class. Quantum Grav. {\bf 15} (1998) 875-905, [gr-qc/9606090];
``III.
Quantum Constraint Algebra and Physical Scalar Product in Quantum General
Relativity", Class. Quantum Grav. {\bf 15} (1998) 1207-1247,
[gr-qc/9705017];
``IV. 2+1 Euclidean Quantum Gravity as a model to test 3+1
Lorentzian Quantum Gravity", Class. Quantum Grav. {\bf 15} (1998) 
1249-1280, [gr-qc/9705018]; 
``V. Quantum Gravity as the Natural Regulator of the Hamiltonian 
Constraint
of Matter Quantum Field Theories",
Class. Quantum Grav. {\bf 15} (1998) 1281-1314, [gr-qc/9705019];
``VI. Quantum Poincar\'e Algebra and a Quantum Positivity of Energy
Theorem for Canonical Quantum Gravity",
Class. Quantum Grav. {\bf 15} (1998) 1463-1485, [gr-qc/9705020];
``Kinematical Hilbert Spaces for Fermionic and
Higgs Quantum Field Theories",
Class. Quantum Grav. {\bf 15} (1998) 1487-1512, [gr-qc/9705021]


\bibitem{8.2} 
A. Ashtekar, J. Lewandowski, D. Marolf, J. Mour\~ao, T.
Thiemann, ``Quantization for diffeomorphism invariant theories
of connections with local degrees of freedom", Journ. Math. Phys.
{\bf 36} (1995) 6456-6493, [gr-qc/9504018]

\bibitem{7.4} H. Sahlmann, ``When do Measures on the Space of Connections
Support the Triad Operators of Loop Quantum Gravity?'', gr-qc/0207112;
``Some Comments on the Representation Theory of the Algebra Underlying
Loop Quantum Gravity'', gr-qc/0207111\\
H. Sahlmann, T. Thiemann, ``On the Superselection Theory of
the Weyl Algebra for Diffeomorphism Invariant Quantum Gauge Theories'',
gr-qc/0302090;
``Irreducibility of the Ashtekar-Isham-Lewandowski Representation'',
gr-qc/0303074\\
A. Okolow, J. Lewandowski, ``Diffeomorphism Covariant
Representations of the Holonomy Flux Algebra'', gr-qc/0302059

\bibitem{7.5} 
A. Ashtekar, C.J. Isham, ``Representations of the Holonomy
Algebras of Gravity and Non-Abelean Gauge Theories",
Class. Quantum Grav. {\bf 9} (1992) 1433, [hep-th/9202053]\\
A. Ashtekar, J. Lewandowski, ``Representation
theory of analytic Holonomy $C^\star$ algebras", in ``Knots and
Quantum Gravity", J. Baez (ed.), Oxford University Press, Oxford 1994





\bibitem{2.1} J. Klauder, ``Universal Procedure for Enforcing Quantum 
Constraints'', Nucl.Phys.B547:397-412,1999, [hep-th/9901010];
``Quantization of Constrined Systems'', Lect. Notes Phys. 
{\bf 572} (2001) 143-182, [hep-th/0003297]\\
A. Kempf, J. Klauder, ``On the Implementation of Constraints through 
Projection Operators'', J. Phys. {\bf A34} (2001) 1019-1036,
[quant-ph/0009072]



\bibitem{6.1} J.E. Marsden, P.R. Chernoff, ``Properties of Infinite
Dimensional Hamiltonian Systems", Lecture Notes in Mathematics,
Springer-Verlag, Berlin, 1974

\bibitem{8.1} T. Thiemann, ``Quantum Spin Dynamics (QSD): VIII.
The Master Constraint'', in preparation


\bibitem{8.3} H. Sahlmann, T. Thiemann, ``Towards the QFT on
Curved Spacetime Limit of QGR. 1. A General Scheme'', [gr-qc/0207030];
``2. A Concrete Implementation", [gr-qc/0207031]

\bibitem{8.5} T. Thiemann, ``Quantum Spin Dynamics (QSD): VII.
Symplectic Structures and Continuum Lattice Formulations of
Gauge Field Theories", Class.Quant.Grav.18:3293-3338,2001,
[hep-th/0005232]; ``Gauge Field Theory Coherent States (GCS): I.
General Properties", Class.Quant.Grav.18:2025-2064,2001, [hep-th/0005233];
``Complexifier Coherent States for Canonical Quantum General Relativity", 
gr-qc/0206037\\
T. Thiemann, O. Winkler, ``Gauge Field Theory Coherent States
(GCS): II. Peakedness Properties", Class.Quant.Grav.18:2561-2636,2001,
[hep-th/0005237]; ``III. Ehrenfest Theorems",
Class. Quantum Grav. {\bf 18} (2001) 4629-4681, [hep-th/0005234];
``IV. Infinite Tensor Product and Thermodynamic Limit",
Class. Quantum Grav. {\bf 18} (2001) 4997-5033, [hep-th/0005235]\\
H. Sahlmann, T. Thiemann, O. Winkler, ``Coherent States for
Canonical Quantum General Relativity and the Infinite Tensor Product
Extension", Nucl.Phys.B606:401-440,2001
[gr-qc/0102038]

\bibitem{RS} M. Reed, B. Simon, ``Functional Analysis'', vol. 1, Academic 
Press, New York, 1980



675-693 (1994) 


\bibitem{B} 
J. Brunnemann, T. Thiemann, ``Simplification of the Spectral Analysis of 
the Volume Operator in Loop Quantum Gravity'', gr-qc/0405060

\bibitem{8.4} E. Witten, Nucl. Phys. {\bf B311} (1988) 46

\bibitem{1.10} D. Giulini, D. Marolf , ``A Uniqueness Theorem for Constraint 
Quantization"
Class. Quant. Grav. 16:2489-2505,1999, [gr-qc/9902045];
``On the Generality of Refined Algebraic
Quantization", Class. Quant. Grav. 16:2479-2488,1999, [gr-qc/9812024]


\bibitem{2.2}  I. M. Gel'fand, N. Ya. Vilenkin, ``Generalized Functions",
            vol. 4, Applications of Harmonic Analysis, Academic Press,
            New York, London, 1964



\end{thebibliography}
\end{document}